\documentclass[twoside]{LCWS11}
\usepackage[latin1]{inputenc}
\usepackage[dvips]{graphicx,color}
\usepackage{epsfig}
\usepackage{wrapfig,rotating,url}
\usepackage{amssymb,amsmath,array}
\usepackage{cite}
\begin{document}
\title{
Recent progress for Linear Collider SM/BSM Higgs/Electroweak
Symmetry Breaking Calculations} 
\author{J\"urgen Reuter$^1$
\thanks{Dedicated to the memory of our friend Uli Baur.}
\vspace{.3cm}\\
DESY Theory Group - \\
Notkestr. 85, D-22603 Hamburg, Germany
}

\maketitle

\begin{abstract}
In this paper I review the calculations (and partially simulations
and theoretical studies) that have been made and published during the
last two to three years focusing on the electroweak symmetry breaking
sector and the Higgs boson(s) within the Standard Model and models
beyond the Standard Model (BSM) at or relevant for either the
International Linear Collider (ILC) or the Compact Linear Collider
(CLIC), commonly abbreviated as Linear Collider (LC).
\end{abstract}

\section{General Remarks}

Most of the work on the electroweak symmetry breaking sector can be
grouped into the following three categories: 1) Precision calculations
for the Higgs mass within the SM and in BSM models, 2) precision
calculations for electroweak (EW) processes relevant to the EW sector,
and 3) Higgs production processes. Each of the categories comprises
studies and calculations within the SM as well as in BSM models. The
first category on precision calculations for the Higgs mass is still
one of the most important issues for the reduction of the theory error
on the Higgs mass prediction which is highly relevant for the Higgs
searches and potential measurements at the Large Hadron Collider
(LHC). This part and also the third part, the precision calculations
for specific Higgs production (and decay) channels are not covered,
mainly because there has not been done much on these subjects in the
past three years, or the results are covered elsewhere. Several topics
that fit in the context of this document have also been covered by
other people during the LCWS conference in Granada like Higgs
production in SUSY decays~\cite{SvHeinemeyerTalk}, composite Higgs
physics~\cite{christophe}, the non-minimal flavor-violating (NMFV)
MSSM~\cite{arana}, precision studies of the 2HDM~\cite{howie}, or also
the Higgs sector of a pure $B-L$ model~\cite{pruna}. 


\section{Precision calculations to SM Higgs and Background Processes} 

\subsection{Electroweak Triboson Production}

Two of the most important processes to check and overconstrain the
sector of electroweak symmetry breaking both at LHC as well as a
future linear collider are triboson production and vector boson
scattering. The processes $e^+ e^- \to WWZ,m ZZZ$ are commonly viewed
as a logical continuation of the physics from $WW$ production at

\begin{figure}
\centerline{
  \includegraphics[width=0.49\columnwidth]{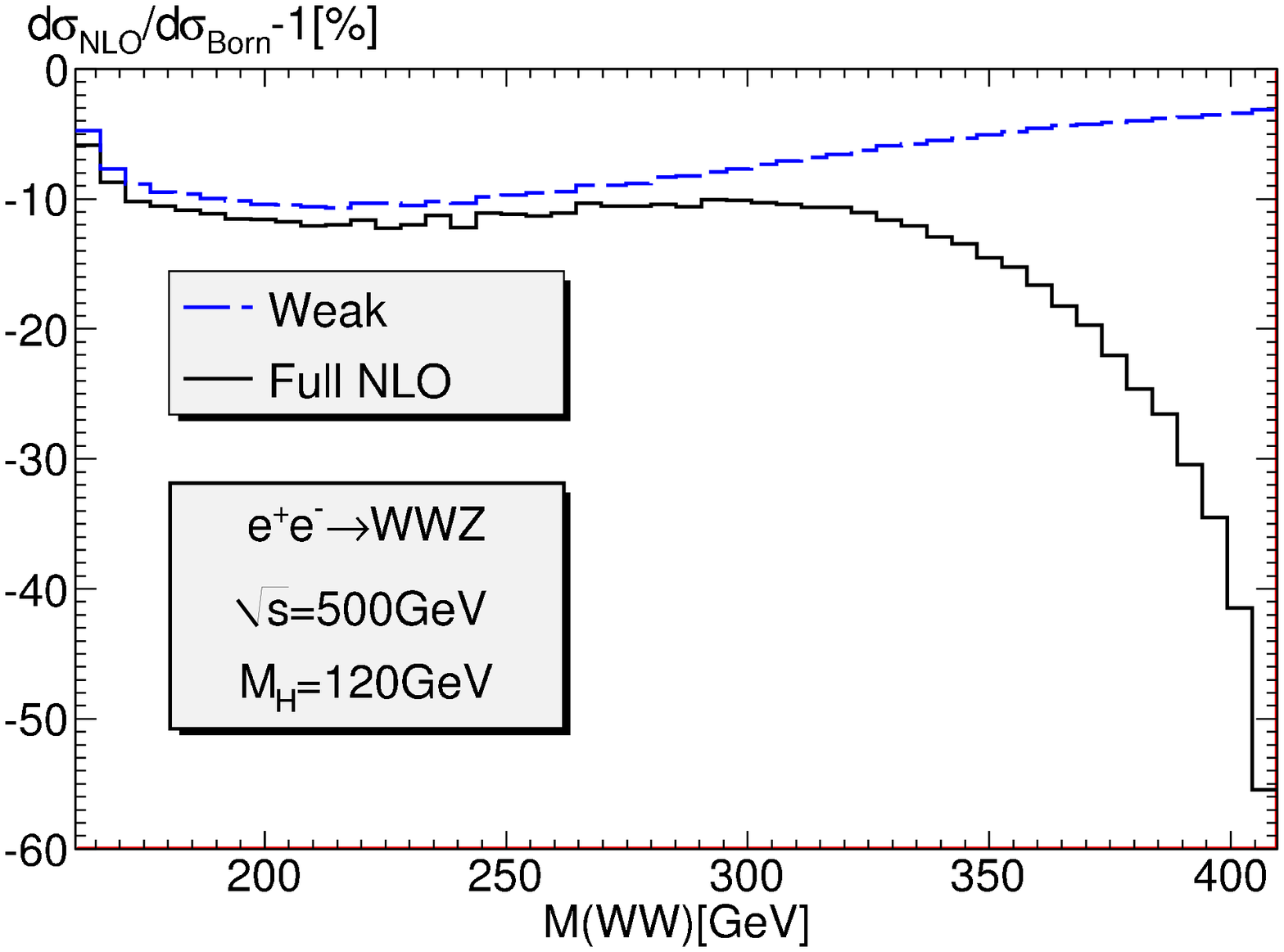} 
  \includegraphics[width=0.49\columnwidth]{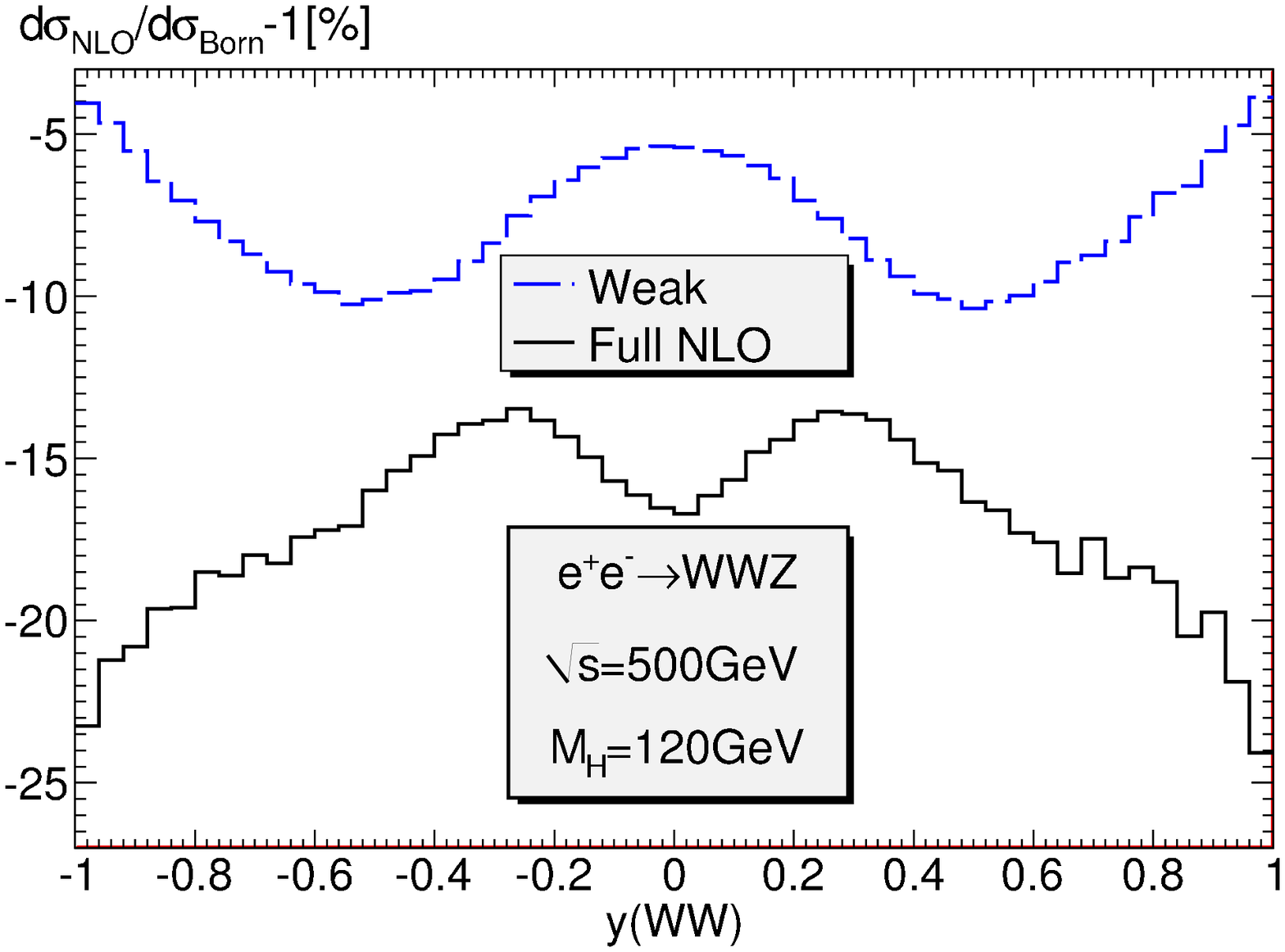}
}
\caption{Differential $M_{WW}$ and $y_{WW}$ distributions for $ee \to
  WWZ$. The black line are the full NLO corrections, while the blue
  line shows the genuine weak corrections.}\label{Fig:wwz_nlo}
\end{figure}

LEP. In Ref~\cite{Beyer:2006hx} it was shown that both 
processes at a 1 TeV ILC allow for a precision measurement of the
quartic gauge couplings (QGC) and a possible determination anomalous
QGCs or constraining them much better than the projected measurement
capabilities at the LHC. The $WWZ$ channel alone enables one to
constrain the parameter space in the $\alpha_{4/5}$ anomalous QGC
parameters, while combining this with the $ZZZ$ measurement shrinks
the confidence level ellipses besides its much lower statistics by a
factor three to four. To determine deviations from the SM QGC with an
integrated luminosity of 1 fb${}^{-1}$ a theoretical precision of at
the per-mil level. There have been three different groups calculating
the NLO corrections to the two triboson production processes $e^+ e^-
\to WWZ, ZZZ$,~\cite{Boudjema:2009pw,JiJuan:2008nn,Wei:2009hq}. 
In the 't Hooft-Feynman gauge used in Ref.~\cite{Boudjema:2009pw} the
NLO corrections comprise 2700 diagrams for the $WWZ$ channel including
109 pentagon diagrams, while the corresponding $ZZZ$ process has
roughly 1800 diagrams containing 64 pentagons. They used two
different independent codes, both based on the
\texttt{FeynArts/FormCalc} package~\cite{Hahn:2000kx,Hahn:1998yk}. The
calculation has been performed in the kinematic regime where $M_{WW}$
is below the Higgs threshold, using the on-shell renormalization
scheme, the Passarino-Veltman tensor
reduction~\cite{Passarino:1978jh},  and the method proposed
in~\cite{Denner:2005nn} to avoid numerical instabilities from
vanishing inverse Gram determinants. They compared the Catani-Seymour
dipole subtraction method~\cite{Catani:1996vz,Catani:2002hc} as well
as a phase-space slicing method to treat the soft-collinear
divergencies, which agree until the soft-collinear approximation
breaks down. Both
groups~\cite{Boudjema:2009pw,JiJuan:2008nn,Wei:2009hq} compared their
results with each other and found agreement. The dominant electroweak
corrections come from QED initial state radiation (ISR), which can be
subtracted to obtain the genuine weak corrections based either on the
Catani-Seymour method or on an experimental extraction from $ZZZ$. The
peak cross section $\sigma_{\text{peak}} (WWZ) \sim 50$ fb is much
larger than the corresponding $\sigma_{\text{peak}} (WWZ) \sim 1.2$
fb. The full EW corrections including QED ISR are negative and amount
to -30 \% of the cross section, while the genuine EW corrections are
of the order of -7 to -18 \%, cf. Fig.~\ref{Fig:wwz_nlo} This
calculation is consistent with estimates from EW double-logarithmic
Sudakov corrections~\cite{Kuhn:1999de}. 

The authors of~\cite{Pasechnik:2010yu} have studied the influence of
finite-width effects in the near-threshold production at the ILC. They
find that using finite widths for EW gauge bosons as well as the Higgs
could alter the lineshape of the process cross sections at the level
of 20-30 per cent. However, a proper treatment using resummation would
be highly desirable here. 

\subsection{Vector Boson Scattering}

\begin{figure}
\centerline{
  \includegraphics[width=0.49\columnwidth]{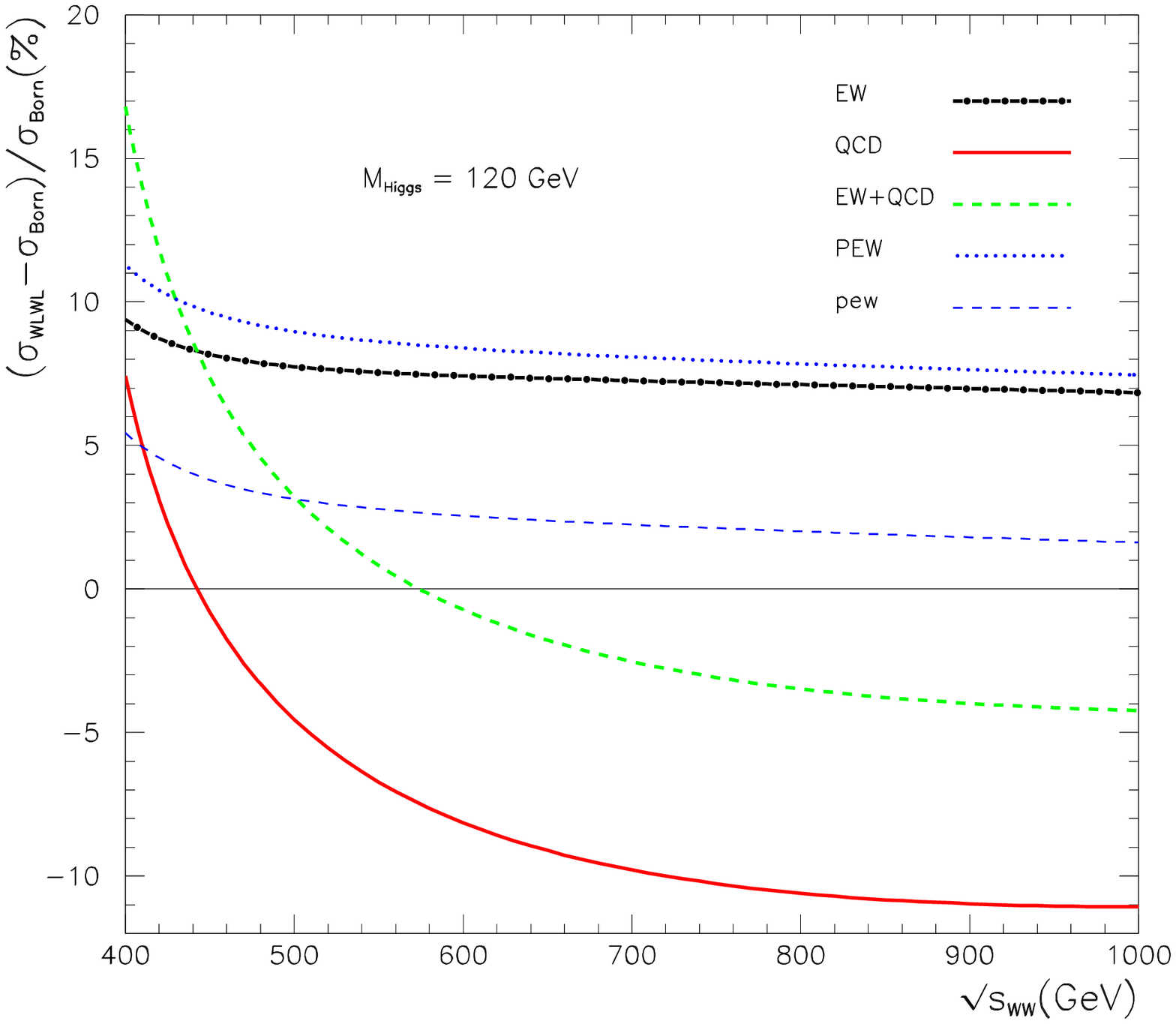} 
  \includegraphics[width=0.49\columnwidth]{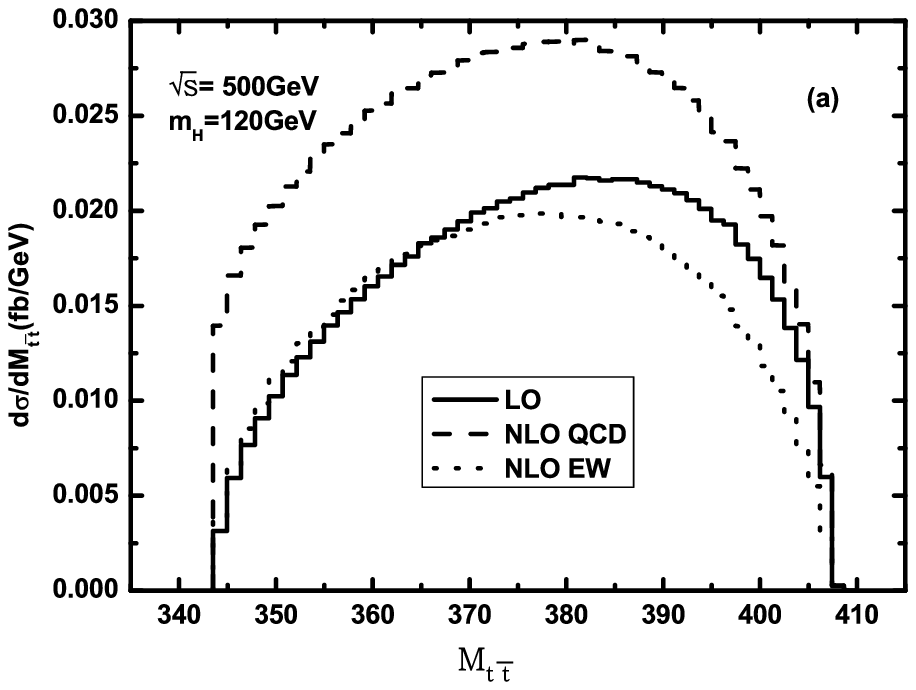} 
}
\caption{\label{vbs_nlo} Left: relative NLO corrections for the
  process $WW\to tt$: 
  the dashed green line are the full corrections, the black line the
  EW and the red the QCD corrections, respectively.
  Right: The $tt$ invariant mass distribution at LO
  (dark line), the dashed line gives the NLO QCD corrections, while
  the dotted one shows the NLO EW corrections.
}
\end{figure}

As has been also shown in~\cite{Beyer:2006hx}, vector boson scattering
is an even more powerful tool to examine and (over)constrain the EWSB
sector (for an overview how to describe resonances in the EW sector
for center-of-mass energies of 1 TeV and larger see
e.g.~\cite{Alboteanu:2008my}). The top Yukawa coupling might play a
special role in the SM as the only coupling of order one, hence, it is
an interesting topic to study vector boson scattering to a top quark
pair. In~\cite{Bouayed:2007rt}, the authors have calculated the QCD
and EW NLO corrections to this scattering process $WW, ZZ \to t\bar t$
in the on-shell scheme. They find that the corrections grow with
rising Higgs mass and that the EW corrections mostly cancel the QCD
corrections such that the total corrections over most of the mass
range are only of the order of $\pm 5$ \%, cf. the left plot in
Fig.~\ref{vbs_nlo}. The 
calculations are for the $2\to 2$ scattering process, and the final
results have been obtained in the effective W approximation (EWA),
i.e. using structure functions for the $W/Z$ as a parton in the
electron. Such a description is known to give rise to deviations of
differential cross sections from irreducible background from $W/Z$ 
bremsstrahlung. This effect can be even larger than the NLO
corrections~\cite{Alboteanu:2008my}. 

\subsection{Top-Antitop Associated Production}

\begin{figure}
\centerline{
  \includegraphics[width=0.7\columnwidth]{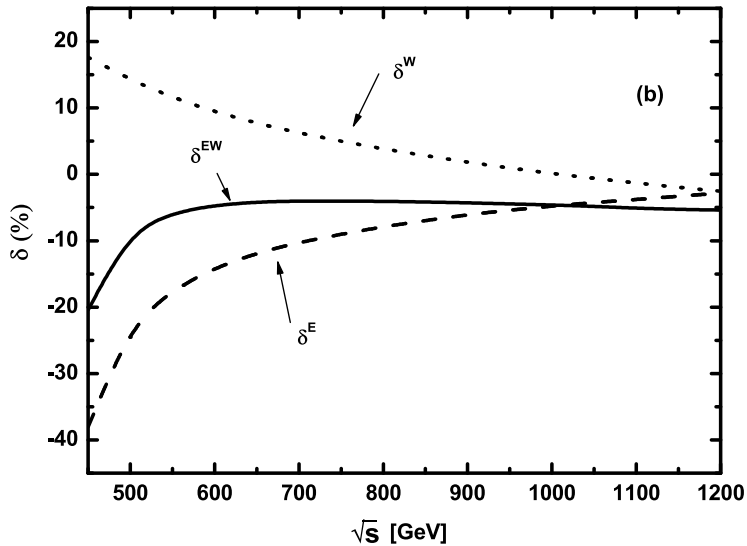} 
}
\caption{\label{ttz} NLO corrections to $ttZ$ production at an ILC:
  The dashed curve shows the electric corrections, the dotted one the
  pure electroweak and the full line the sum of the two.
}
\end{figure}

At ILC or CLIC, $t\bar th$ associated production is an important
process to measure precisely the properties of the Higgs boson after
it has been discovered, specifically its Yukawa coupling to the top
quark. For this process and a light Higgs boson which predominantly
decays into bottom quarks the process $e^+ e^- \to t\bar t Z$ is one
of the most severe backgrounds. Hence, it is important to know this
background at NLO. The QCD and EW NLO corrections have been calculated
in the paper~\cite{Lei:2008jx} based on
\texttt{FeynArts/FormCalc}~\cite{Hahn:2000kx,Hahn:1998yk}. The
on-shell scheme has been used and a trivial CKM matrix was assumed
throughout the calculation which has been performed for a fixed Higgs
mass of 120 GeV. Lei et al. used photon and gluon masses,
respectively, to regulate infrared singularities. For this process, as
expected, the QCD corrections dominate and are positive, $\sim 40$ \%,
while the EW corrections are negative of size $\sim - 4-8$ \%, see
the right plot in Fig.~\ref{vbs_nlo}. The relative size of purely
electromagnetic as well as purely weak corrections are shown in
Fig.~\ref{ttz}.

\subsection{Charged Higgs production}

Within the Two-Higgs-doublet model, either the generic one, or in the
context of the MSSM, charged Higgs pair production might not be in the
kinematical of a 1 TeV ILC. It is nevertheless worthwhile to study the
production where only one of the two is on-shell and then goes into a
top-bottom pair, hence the process: $e^+ e^- \to t \bar b H^-$. This
is specifically important in the high Higgs mass region and in the
decoupling limit. The SM QCD NLO corrections for that process have
been calculated in~\cite{Kniehl:2002zz}. Quite recently, the SUSY QCD 
NLO corrections for that process have been calculated
by~\cite{Kniehl:2010ea}. They found that the corrections are enhanced
in the parameter region of large $\tan\beta$ which can as usual be
accounted for by a resummed bottom Yukawa coupling. The residual SUSY
QCD corrections are of the order of $-10$ to $-15$ \%. In addition,
the authors checked a completely analytical calculation by a method
using a numerical evaluation of both loop and phase-space integrals
based on Bernstein-Tkachov~\cite{Bernstein,Tkachov:1996wh}. This
method which has been used before only for up to $2\to 2$
processes~\cite{Passarino:2010qk}, is here for the first time applied
to a $2\to 3$ kinematics. The authors found no particular gain in
speed/performance or a simpler treatment compared to the fully
analytic approach.    

\subsection{Trilinear Higgs coupling}

The trilinear Higgs coupling is the main recent focus on
investigations in SM Higgs physics, as it might give the only possible
handle to the Higgs potential itself, which is the true trigger for
the EWSB. Most of the relevant investigations about measuring the
triple Higgs coupling at the ILC can be found
in~\cite{Djouadi:1999gv,Muhlleitner:2000jj}. 
More recently, in~\cite{GutierrezRodriguez:2008nk} the processes $e^+ e^-
\to HH b\bar b, HHt\bar t$ have been studied with the main purpose to
look for possible interference effects from the continuum production
and the diagram containing the trilinear Higgs coupling. As expected,
the interferences are tiny, of the order 3 \%, for these processes. 
A very extensive investigation of the question whether and how well
the trilinear Higgs coupling will be measurable at ILC has been done
by Baur~\cite{Baur:2009uw}.

\begin{figure}
\begin{center}
  \includegraphics[width=0.65\columnwidth]{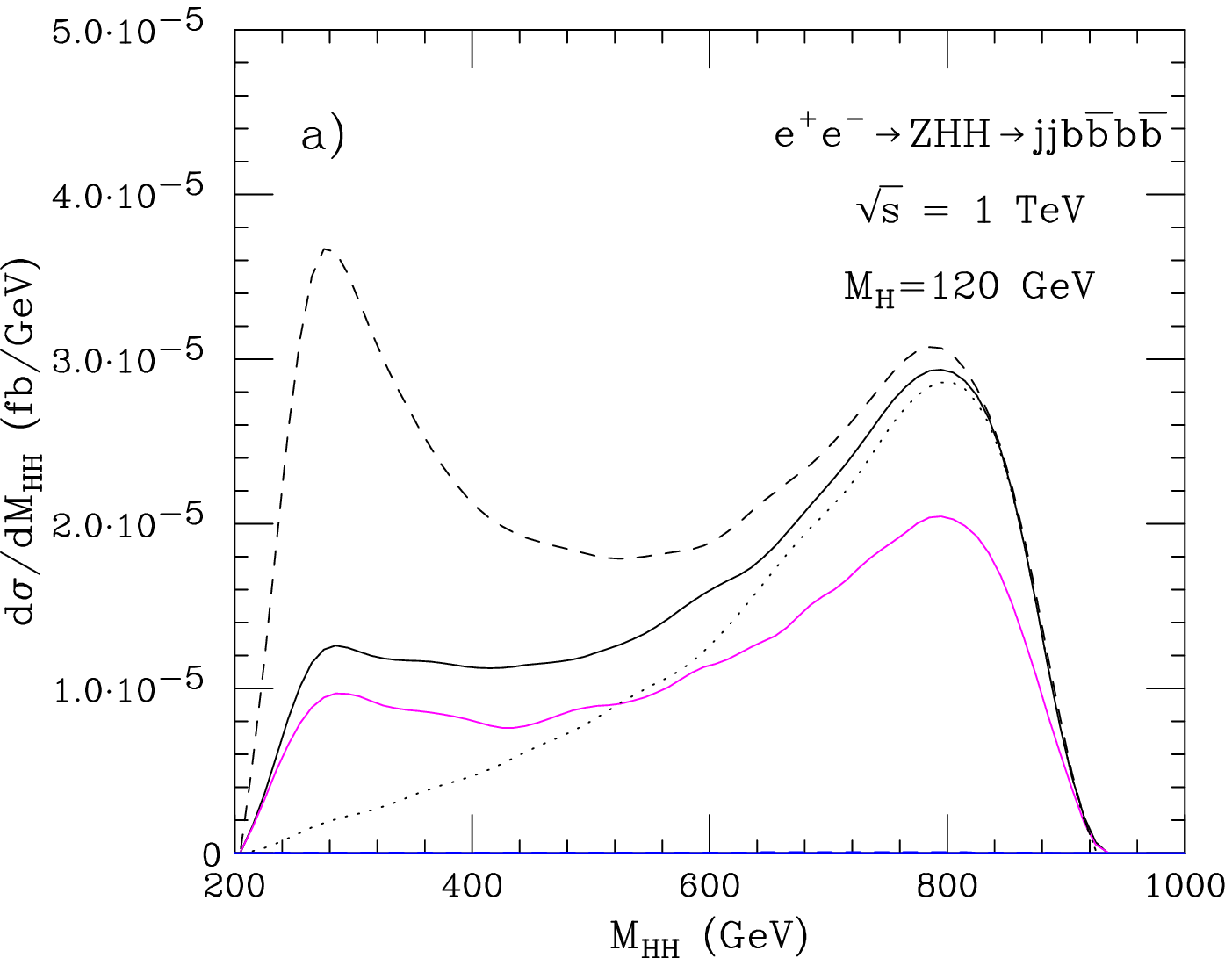} \\
  \includegraphics[width=0.65\columnwidth]{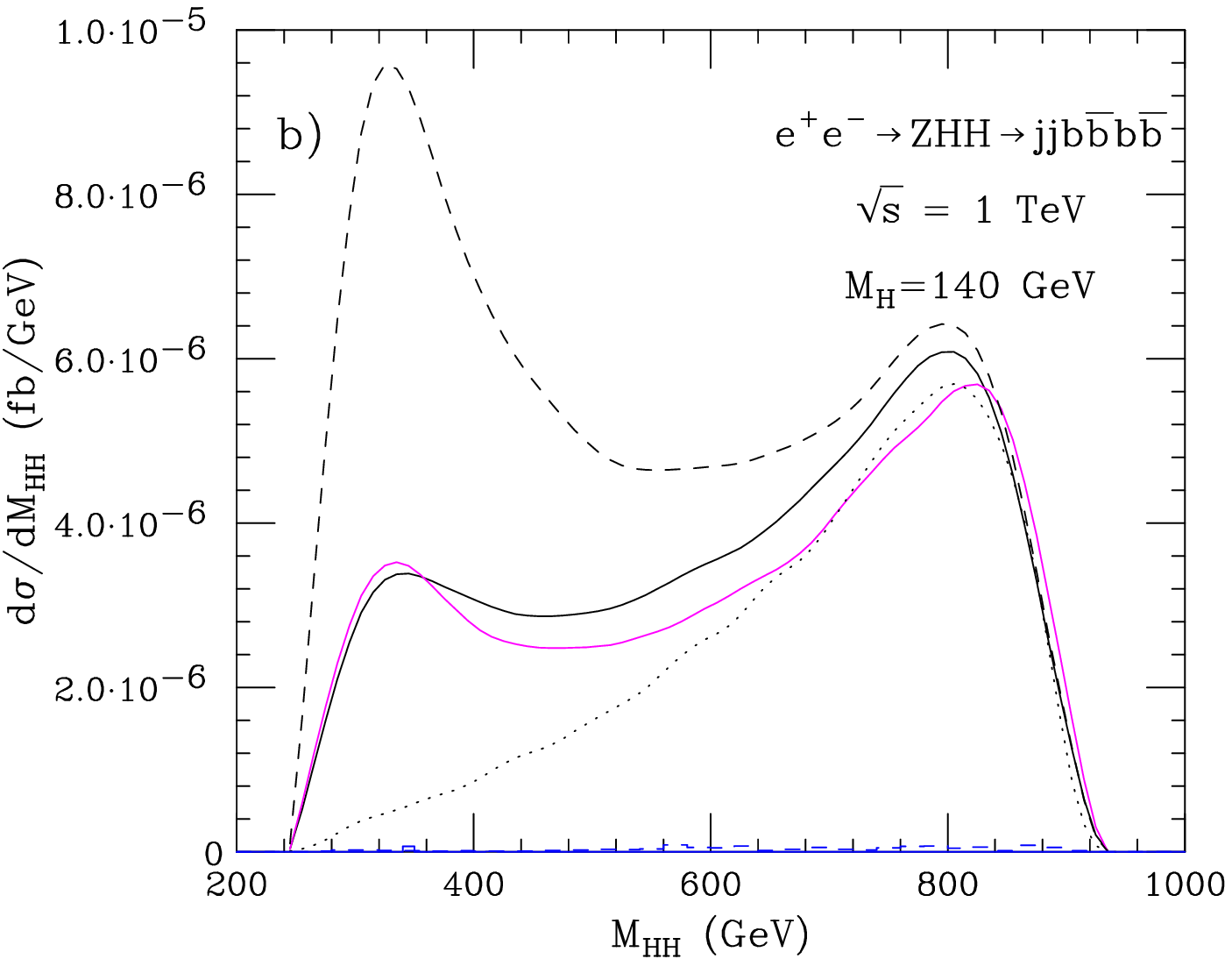} 
\end{center}
\caption{\label{hhh} $HH$ invariant mass in the fully reconstructible 
final state $jjbbcc$ for Higgs masses of 120 GeV (upper) and 140 GeV
(lower). The dashed line is for enhancing the trilinear coupling by a
factor of 2, the dotted line is setting it to zero. The blue line
shows the factorized process without interferences. 
}
\end{figure}

Baur found that the $WW$ fusion process with the final state
$\nu\bar\nu HH$ dominates over the Higgsstrahlung process with final
state $ZHH$ (note that the first is partially also contained in the
second one regarding the invisible $Z$ decay, but not in the
kinematical region considered here). The paper simulates the full
final states taking all interferences into account using the
\texttt{WHIZARD}
generator~\cite{Kilian:2007gr,Moretti:2001zz}. Off-shell and
interference effects are known to be generically crucial in
electroweak production at an ILC, specifically if heavier Higgses are
to be extracted via cut-based analyses from their SM
backgrounds~\cite{Hagiwara:2005wg}. Despite this, these effects are
not overly important for that particular process here. The main
conclusion of the paper is that the trilinear Higgs coupling
$\lambda_{HHH}$ can be measured for a Higgs mass in the range of
120-180 GeV at a 1 TeV ILC with a precision from
20-80 \%, while a 3 TeV CLIC could even achieve a precision of 10-20
\%. The result although has to be taken with a small grain of salt as
neither ISR nor beamstrahlung have been included in the study; but
they both can have quite large effects on such signal-to-background
investigations (cf. e.g.~\cite{Hagiwara:2005wg}). The largest
background to the signal comes from the $jjbbcc$ final state with two
light jets, two $b$ jets and two charm jets with charm mistagged as a
bottom quark. The di-Higgs invariant mass in the fully reconstructible
final state $jjbbcc$ is shown for two different Higgs masses (120 and
140 GeV) in Fig.~\ref{hhh}.



\section{EW processes in BSM models}

Many studies on BSM models have been done, historically with a strong
focus or even bias to supersymmetric models like the MSSM (cf. 
e.g.~\cite{Accomando:1997wt,AguilarSaavedra:2001rg,AguilarSaavedra:2005pw}. 
Here, I will not cover any EW calculations or studies related to
supersymmetric theories, some work presented during this conference
can be found in~\cite{SvHeinemeyerTalk}. The investigations summarized
in this report deal with a SM with a fourth generation, technicolor
and topcolor-assisted technicolor, Little Higgs models as well as
twin-Higgs models. 

\subsection{SM with a fourth generation}

\begin{figure}
\centerline{
  \includegraphics[width=0.49\columnwidth]{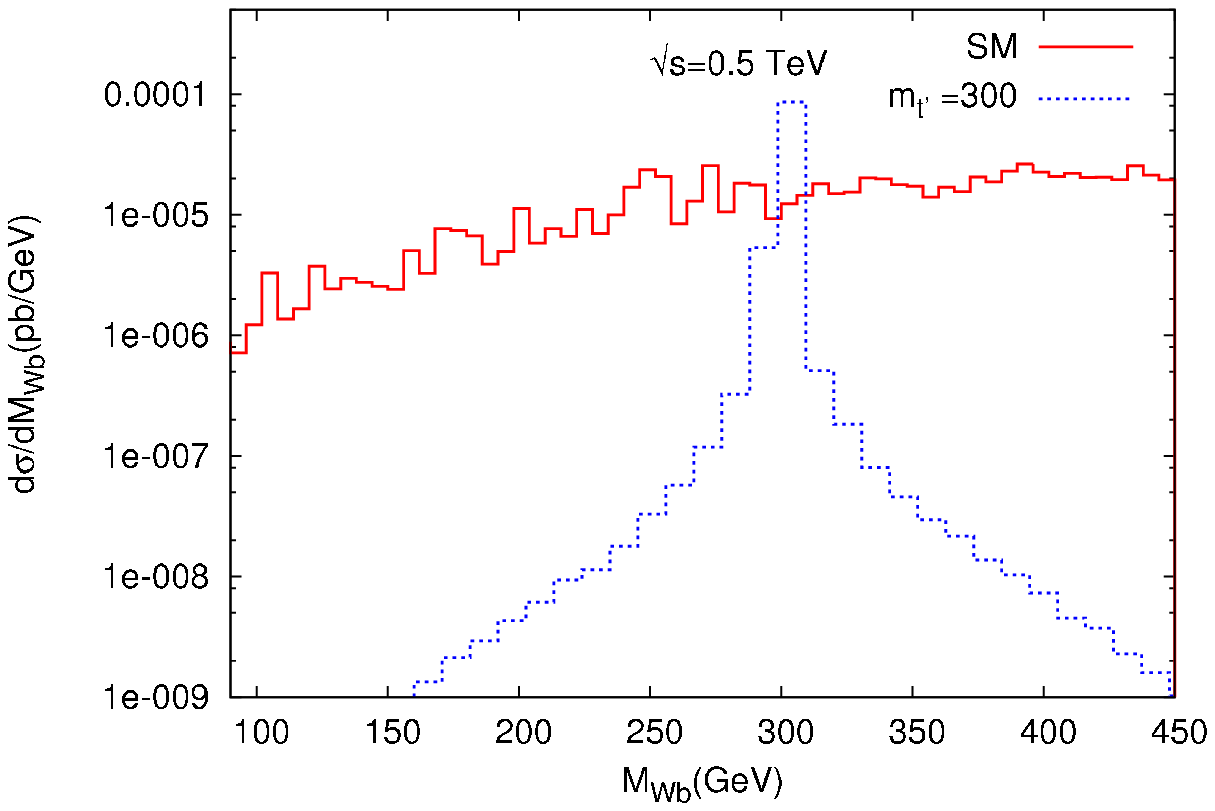} 
  \includegraphics[width=0.49\columnwidth]{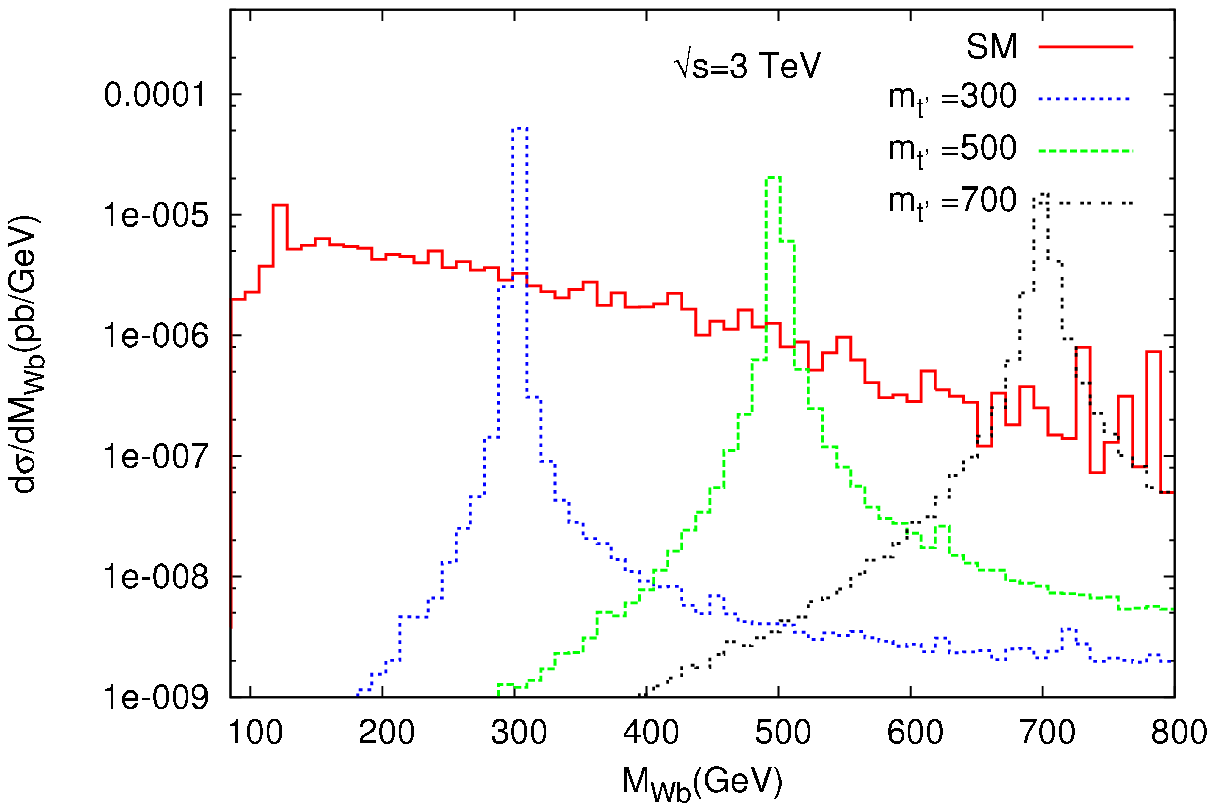} 
}
\caption{\label{4gen} The $Wb$ invariant mass for center-of-mass
  energies of 500 GeV (left) and 3 TeV (right). The red line shows the
  SM background, the peaks are the $t'$ signals for different masses. 
}
\end{figure}

A fourth SM family is already heavily constrained by LHC data, not so
much by the direct searches but by the Higgs search. However, such a
possibility is still not completely ruled out. In the
paper~\cite{Senol:2011nm} special type of dimension six interactions
between the light SM quarks, the 4th generation up-type quark ($t'$)
and the SM gauge bosons are introduced which resemble magnetic
moment-type interactions. Such operators like $\mathcal{L} = \sum_u
\frac{\kappa}{\Lambda} \overline{t'} \sigma_{\mu\nu} q_u F^{\mu\nu}$
could even dominate the chiral SM interactions, and are hence easy,
but also worthwile to study. These couplings allow for a single
production of the $t'$ states in $e^+e^- \to t'q \to Wbq$. The authors
study the detectability at ILC in these final states as a function of
the $t'$ mass, shown in Fig.~\ref{4gen}, where the $Wb$ invariant mass
is shown for a 500 GeV ILC and a 3 TeV CLIC for different $t'$ masses.  


\subsection{Technicolor and Top-Color assisted Technicolor}

Technicolor is a model where there are new matter constituents
(techniquarks) and the corresponding force carriers (technigluons) of
new strong dynamics at the TeV
scale~\cite{Weinberg:1975gm,Susskind:1978ms}. These models are having
difficulties to be reconciled with the electroweak precision
observables, with flavor, and the generation of fermion
masses. However, they gained a big revival in the context of dual
models in the sense of the adS/CFT correspondence. Topcolor, on the
other side, is a model where there is a new strong interaction of only
the top quark which makes it condense and triggers EW symmetry
breaking~\cite{Pendleton:1980as,Hill:1980sq}. Topcolor-Assisted
Technicolor (TC2) is a mixture of both (extended) technicolor as well
as topcolor, where one is forced to introduce a vacuum ``tilted'' in
the $U(1)$ charges to avoid phenomenologically catastrophical $b\bar
b$ condensates (for more details cf.~\cite{Hill:2002ap}). 

In these models there are emergent top pions with masses naturally of
the order of the top quark, $m_\pi \sim m_t$, a top-pion decay
constant of $f_\pi \sim 60$ GeV, and a corresponding ``Yukawa''
coupling $g_{tb\pi} \sim m_t/sqrt{2} f_\pi \sim 2.5$. So basically,
these models are like a 2HDM in the decoupling limit, with the
pseudoscalar top-pions and a scalar called the top-Higgs as decoupled
states. In~\cite{Han:2011jv} the authors calculated the pair
production in the processes $\pi_t\pi_t$ and $\pi_t h_t$ both for the
LHC as well as for the ILC. The indirect bounds together with the
searches from Tevatron (which had almost never been able to see these
states) yield a lower bound like $m_\pi, m_h \sim 220$ GeV. Even with
present LHC data presumably only the low-mass region is left
non-excluded, but searches for these states at the LHC are
difficult. The authors calculated the tree-level cross sections as
well as the NLO vertex correections using the \texttt{LoopTools}
package~\cite{Hahn:1998yk}. They found small K-factors of 1.05. The
cross section for $e^+ e^- \to \pi_t \pi_t$ at a 1.5 TeV linear
collider is of the order 20 fb. 

For the two processes $e^+e^- \to \pi_t \pi_t$ and $e^+e^- \to \pi_t
h_t$ there is a discrepancy with two earlier NLO calculations,
Ref.~\cite{Wang:2005qt} and~\cite{Qiao:2008rn}. The main backgrounds
for these processes are SM triboson production which can however
effectively be suppressed by a cut-based analysis. 

Another paper~\cite{Liu:2011pu} calculates the pair production cross
sections of topcolor pions in photon-induced processes, both at the
LHC and at the ILC. Together with the tree-level result they also
calculate the leading NLO contribution. For the photon-induced
toppion pair production, $e^+ e^- \to \gamma\gamma \to \pi_t \pi_t$
the K-factor is again 1.05, while the cross section is of the order
of one pb at a 1.5 TeV linear collider. For the toppion/top-Higgs pair
production the corresponding cross section is $\lesssim 10$ fb. The
main conclusions from here are that the ILC (at least in the 1.5 TeV
version) can detect top-pions, but cannot compete with the LHC
detection capabilities. For all of these papers mentioned here, an
investigation of a realistic ILC environment together with ISR and
beamstrahlung is missing as well as a more thorough study of the ILC
capability as a function of the parameter spaces of the TC
models. Note that there was also another independent NLO calculation
for $\gamma\gamma \to \pi_t^+ \pi^-_t$ in Ref.~\cite{Han:2011jv}. 

Further calculations dealt with the process $e^+ e^- \to W^+ \mu_t^-
\pi^0_t$~\cite{Liu:2008cb}, where the cross sections are only in the
range of a few femtobarns. The main background is $ttbcW$ which is
very complicated final state with high multiplicity. Such a signal
will be quite difficult to dig out of the background. Also note that
such cross sections are not reliable any more for center-of-mass
energies $\sqrt{s} \gtrsim 1$ TeV, as in the case of non-unitarized
multi-pion scattering. Ref.~\cite{Liu:2008cb} also studied the
associated production $e^+e^- \to Z\pi_t^+\pi_t^-$. Again, the cross
sections are of the order of roughly a femtobarn and only marginally
visible. All these calculations have also been performed for the case
of photon-induced processes, $ee \to \gamma\gamma \to X$. Another
publication~\cite{Huang:2010vh} also included production of a
top-Higgs in association with top quarks in gamma-induced collisions,
$e^+e^- \to \gamma\gamma \to t\bar t h_t$. 


\subsection{Little Higgs models}

\begin{figure}
\centerline{
  \includegraphics[width=0.8\columnwidth]{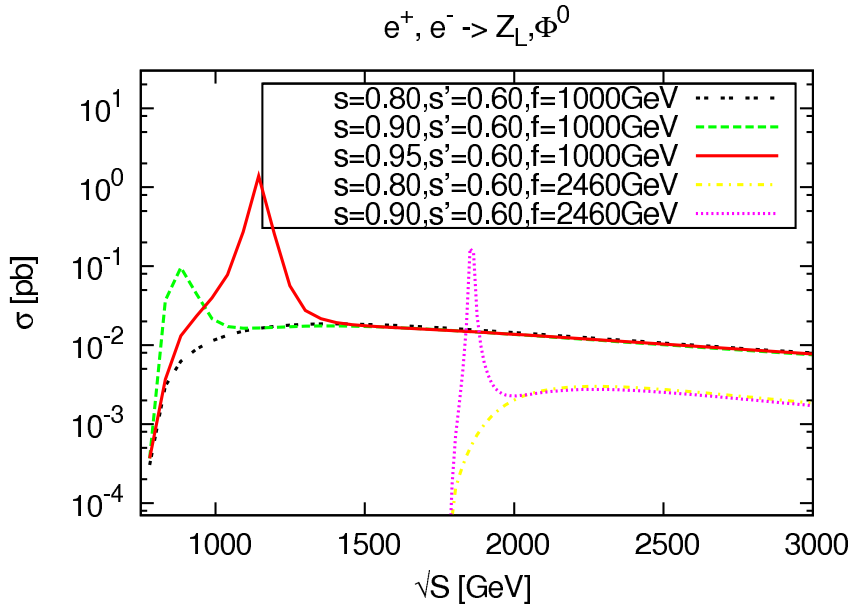} 
}
\caption{\label{lhm_zh} The cross section for the pair production of
  $Z\Phi$ in the Littlest Higgs model as a function of the
  center-of-mass, the two mixing angles in the gauge sector as well as
  the model scale, $f$.
}
\end{figure}

Little Higgs
models~\cite{ArkaniHamed:2002qx,ArkaniHamed:2002qy,ArkaniHamed:2002pa}
are a variant of strongly interacting models that are in better
agreement with EW precision data, as they manage to have a weakly
interacting sector at the TeV scale. These models have as generic
properties an extended global symmetry together with an extended
scalar sector compared to the SM, an extended gauge symmetry and hence
new heavy vectors ($Z'$, $W'$), as well as new heavy fermions,
especially a heavy top quark $T$. Some of these models have a discrete
symmetry, called $T$-parity in order to further ameliorate EW precision
observables. This allows then for dark matter. Generically, there are
two types of models, the product group model, where the scalars are in
an irreducible representation (irrep), while the weak gauge group is
a product group, and the simple group models, where the weak gauge
group is simple and the scalar modes are in a product group
representation. The most important model of the first class is the
Littlest Higgs~\cite{ArkaniHamed:2002qy}, and the Simplest Little
Higgs~\cite{Schmaltz:2004de} of the second class, respectively. 
In the Littlest Higgs, the essential parameters of these models are
the intermediate scale which sets the mass for the additional weakly
interacting states as well as the Higgs triplet vacuum expectation
value (vev). Constraints from EW precision observables demand $f
\gtrsim 3-4$ TeV, and $v' \lesssim 10^{-2}$
GeV~\cite{Csaki:2002qg,Hewett:2002px,Kilian:2003xt}. As the Goldstone
bosons from the complex Higgs triplet of that model are difficult to
see at the LHC, Ref.~\cite{Cagil:2010mk}. The cross section largely
suffers from phase space as well as the decoupling limit,
Fig.~\ref{lhm_zh}. The authors 
focus on the fermiophobic limit which is favored by EW precision
observables. The main SM background is $Ztt$, but no realistic
detectability study has been done (yet). 

In many of these models, there are global $U(1)$ symmetries rather
naturally which lead to single light pseudoscalar Goldstone
bosons~\cite{Kilian:2004pp}, which could serve as a discriminator
between different Little Higgs model classes~\cite{Kilian:2006eh}. 
Recently, the top quark-associated production of such a pseudo-axion
has been revisited for the Simplest Little Higgs with
$T$-parity~\cite{Han:2011zz}. The cross sections are compatible with
the corresponding $ee \to tth$ cross sections from the SM or the MSSM,
namely roughly 1 fb. The authors also added the
$tt\eta$ production from $\gamma\gamma$-induced
interactions. Generically, all these cross sections are not too
promising but they add an additional source for measurements within
the Simplest Little Higgs.  


\subsection{Twin-Higgs models}

Twin-Higgs models are extensions of the SM by a discrete (parity)
symmetry. This could be either a mirror symmetry~\cite{Lee:1956qn}, or
a discrete left-right exchange
symmetry~\cite{Pati:1974yy,Mohapatra:1974hk,Mohapatra:1974gc}. These
models are similar to Little Higgs as there are Goldstone bosons
arising from the breaking of a large(r) global symmetry. The parity
doubling is responsible for the cancellation of the quadratic
divergencies. The special thing about these models is that they have
no new colored states, hence they are notoriously difficult to
discover at the LHC. It is only the Higgs which communicates to the
mirror sector. Drell-Yan production is possible but suffers from large
backgrounds. In the left-right symmetric twin-Higgs
model (LRTH)~\cite{Chacko:2005un}  there is a $U(4)_1 \times U(4)_2$
global symmetry, and a gauge left-right symmetry, $SU(2)_L \times
SU(2)_R \times U(1)_{B-L}$. The Higgs spectrum consists of the
standard scalar Higgs, a charged Higgs $\phi^\pm$, a pseudoscalar
$\phi^0$ and a heavy doublet $(h_1^+, h_2^0)$. In
Ref.~\cite{Liu:2011zzi} the pair production of $h\phi^0$ has been
calculated in the LRTH. The cross sections are marginal and even close
to threshold not larger than one fb, but can give rise to $Z'$
resonances at CLIC in the range above 1 TeV. As the predominant decay
is $\phi^0 \to b \bar b$, these states are only limited by the
detector resolution. Another publication estimated the corrections to
(multiple) Higgsstrahlung $e^+e^- \to ZH, ZHH$ in the LR Twin-Higgs
model,~\cite{Liu:2008zzb}. As has been studied in the context of
Little Higgs models, extensions of the EW Higgs sector can drastically
enhance these processes~\cite{Kilian:2006eh}. The enhancement auf
these cross sections makes the processes available as discovery modes
at an ILC or CLIC. 


\section{Summary and Conclusions}

Screening the literature and work on electroweak and Higgs physics
(precision) calculations of the last couple of years with emphasis on
a Future Linear Collider shows that we are well prepared for the
physics at such a machine. Basically all signal and most background
processes are known at next-to-leading order, while some of the open
processes like vector boson scattering and triboson production have
been calculated at NLO recently. These processes are the cornerstone
of the high-luminosity and/or high-energy physics program, and they
are at the heart of EW symmetry breaking. One of the final open tasks
after a possible discovery of a SM-like Higgs boson and its precision
taxonomy at LHC and ILC/CLIC will be the mapping out of the Higgs
potential to find out whether this is indeed as given in the SM or has
some deeper mechanism leading to its generation. 

Concerning beyond the Standard Model (BSM) physics, the focus in the
recent years was on EW symmetry breaking-related processes in mostly
strongly interacting models like Little Higgs, technicolor, topcolor
and twin-Higgs models. Specifically multiple production of scalar
particles as well as production in association with top quarks or the
EW gauge bosons have been the driving forces of the
investigations. Generically, one can say that guidance is needed from
the Higgs-sector measurements and the high-energy phase of LHC to know
in which direction to turn for a future ILC/CLIC.


\section{Acknowledgments}

I would like to thank the organizers for a wonderful and perfect
conference at one of the most fascinating places in the world.


\begin{footnotesize}


\end{footnotesize}


\end{document}